\documentclass{article}
\usepackage[utf8]{inputenc}

\author{Sawsan Daws$^1$ and David R. Andersen$^2$}
\date{Dated: July 22, 2021}

\usepackage{amsmath}
\usepackage{mathtools}
\usepackage[colorlinks, citecolor=blue, urlcolor=blu, linkcolor=blue]{hyperref} % for references
\usepackage{graphicx} %for images
\usepackage{geometry} 
\usepackage{nccmath}
\usepackage[numbers, sort&compress]{natbib}
\usepackage{authblk}

\begin{document}
\author{Sawsan Daws\thanks{sdaws@uiowa.edu} and David R. Andersen\thanks{k0rx@uiowa.edu}}
\affil{Department of Physics and Astronomy and Department of Electrical and Computer Engineering The University of Iowa, Iowa City, IA 52242, USA}

\title{Third-order terahertz optical response of graphene in the presence of Rabi Oscillations}
\maketitle{}

\textbf{Abstract}\\
Graphene has been shown to exhibit a nonlinear response due to its unique band structure. In this paper, we study the terahertz (THz) response metallic armchair graphene nanoribbons, specifically current density and Rabi oscillations beyond the semiclassical Boltzman model. We performed quantum mathematical modeling by first finding a solution to the unperturbed Hamiltonian for a single Fermion in the dipole gauge and then applying a polarized, THz electrical field. After writing the solution in terms of the four eigenstates of the Dirac system, we numerically calculated the $x$ and $y$ components of the induced current density resulting from applying the terahertz electrical field. Due to the inclusion of the Rabi Oscillation in our calculation of the optical response, we predict both odd and even harmonics, as well as continuum oscillations of the power density spectrum in the THz regime.  Lastly, we show a rapid decay of the power harmonics.


\section{Introduction}

Graphene is a single layer of carbon atoms in a two-dimensional hexagonal arrangement. Its low-energy excitations are massless and chiral, leading the formation of Dirac fermions[1]. These Diract fermions move with speed 300 times slower than the speed of light c, making the atypical characteristics of the quantum electrodynamics appear in graphene at a much slower speed [2]. Other unique properties of graphene have also drawn enormous attention including the ultrahigh electron carrier mobility [3], optical visualization regardless of it being a single-atom thick substance [4], high electrical transmittance allowing for advanced performance in optical applications [5] linear dispersion properties leading to nonlinear optical response in the terahertz frequencies [6]. A fundamental step in studying the nonlinear optical response of graphene is the investigation of the behavior of the Fourier harmonic response due to current density, including the Rabi Oscillations[7-11]

In his work based on the Boltzman theory[12], Mikhailov showed that the current density in graphene produces only odd harmonic, nonlinear electromagnetic excitation response that is due to the Dirac cone structure in the hexagonal-shaped Brillouin zone of graphene [13]. In a more recent work, Lee et. al. [7] illustrated that even in the presence of the centrosymmetry in graphene atomic structure, the nonlinear optical response is not restricted odd harmonic spectra when taking into consideration the dynamics of Rabi Oscillation in the calculations for the current response. Based on this work, Jin et. al. [14] have also shown that the nonlinear optical response of LCS lattice in graphene – to the second neighbor atoms- is not limited to odd harmonic spectra when accounting for the large contribution of Rabi oscillations in the current response.

In this work, we study the Rabi frequency behavior and harmonic spectra of massless 2D Dirac fermions of metallic armchair graphene nanoribbons (acGNR) due to various terahertz electric field amplitudes. Here, we dive deeper into analyzing the response of the fermions by going beyond the semi-classical Boltzman model to the quantum model in the dipole gauge. The lattice structure of acGNR in study consists of three-level system formed from the nearest-neighboring atoms surrounding the main two atom in the hexagonal unit cell of graphene, with these two atoms situated in the corners of the unit cell. The description of the relationship between A and B atoms is derived using tight-binding model by Brey and Fertig [15].

This paper is organized as follows. In Sec. 2 We find solution to the TDDE in the dipole gauge using the eigenstates of the Hamiltonian matrix of the metallic acGNR, then we derive four equations of motions after applying the incident THz electrical field and examining the relationship between those equations of motion. Lastly, in section 2.3 we find expressions for the x and y component of the induced current density as a response due to the applied electric field.
In Sec. 3 we numerically calculate the induced current density for two different light pulse values and illustrate the harmonic effect caused by the Rabi oscillation. Finally, conclusions of the study are given in Sec.4.

\section{Theoretical Model}
\subsection{Eigenstates of Unperturbed Hamiltonian}
Graphene atoms form a 2D hexagonal lattice of covalently bonded carbon atoms. The unit cell may be regarded as a triangle with two basis atoms, labeled A and B. At low energies, the electron and hole bands behave linearly near the Dirac points $\mathbf{K}=\frac{2\pi}{a_0}(\frac{1}{3},\frac{1}{\sqrt{3}})$ and $\mathbf{K'}=\frac{2\pi}{a_0}(-\frac{1}{3},\frac{1}{\sqrt{3}})$ in the Brillouin zone. Here, $a_0$ is the triangular parameter for the unit cell structure, $a_0=\sqrt{3}a_{cc}$ and $a_{cc}$ is the nearest neighbor distance for carbon atoms, $a_{cc}=1.42\mathring{A}$.

The unperturbed Hamiltonian for a single Dirac fermion in the acGNR is described by Wang et al.[16] as:
\begin{equation}
{\bf H_0}=
\begin{bmatrix}
H_{0,K} & 0\\
0 &H_{0,K'}
\end{bmatrix}
\end{equation}
Where 
\begin{equation}
= \hbar v_F
\begin{bmatrix}
0 & k_x-ik_y & 0 & 0 \\
k_x+ik_y & 0 & 0 & 0 \\
0 & 0 & 0 & -k_x-ik_y \\
0 & 0 & -k_x+ik_y & 0 
\end{bmatrix}
\end{equation}
Where $v_F=c/300$ is the Fermi velocity, $k_x$ and $k_y$ are the perturbation components of $\mathbf{k}$ of the Dirac point. The four eigenstates equations of this Hamiltonian follow the general form: 
$$\psi_\mathbf{k}^{(i)} (\mathbf{r})=\frac{1}{2 \pi}\mathbf{u}^{(i)}e^{(\mathbf{ik . r)}}$$ with corresponding eigenvalues 
$\lambda_k^{(1)}= - \hbar v_F \sqrt{k_x^2 +k_y^2},
\lambda_k^{(2)}= - \hbar v_F \sqrt{k_x^2 +k_y^2},
\lambda_k^{(3)}=\hbar v_F \sqrt{k_x^2 +k_y^2},$ and lastly\\ $\lambda_k^{(4)}=\hbar v_F \sqrt{k_x^2 +k_y^2}$. 
We note here the mirror symmetries in this Dirac Fermion Hamiltonian, where the first and second eigenvalues are identical, the third and fourth are identical, and these groups are opposite of each other. This is due to honeycomb nature of the graphene crystal lattice where the $\mathbf{K}$ and $\mathbf{K}$' are located in the Dirac cone site. 

Lastly in this section, we introduce the normalized eigenstates corresponding to the unperturbed Hamiltonian to use them in finding a solution to the Hamiltonian in the dipole gauge:
 
\begin{equation}
{\bf u_k(+,\uparrow)}^{(1)}= \frac{1}{\sqrt{2}}
\begin{bmatrix}
0\\
0\\
\frac{k}{k_x - ik_y}\\
1
\end{bmatrix} 
,
{\bf u_k}(+,\downarrow)^{(2)}= \frac{1}{\sqrt{2}}
\begin{bmatrix}
\frac{-k}{k_x + ik_y}\\
1\\
0\\
0\\
\end{bmatrix}
\end{equation}

\begin{equation}
{\bf u_k}(-,\uparrow)^{(3)}= \frac{1}{\sqrt{2}}
\begin{bmatrix}
0\\
0\\
\frac{-k}{k_x - ik_y}\\
1
\end{bmatrix} 
,
{\bf u_k}(-,\downarrow)^{(4)}= \frac{1}{\sqrt{2}}
\begin{bmatrix}
\frac{k}{k_x + ik_y}\\
1\\
0\\
0\\
\end{bmatrix}
\end{equation}
These eigenstates satisfy the orthonormality condition: $\int d^2 \mathbf{r} [ \boldsymbol{\psi_{k'}}^{(s)} (\mathbf{r})]^\dagger \boldsymbol{\psi_k}^{(s)} (\mathbf{r})= \delta (\mathbf{k}-\mathbf{k'}) \delta_{ss'}$ with $s'=\pm $ is the pseudospin index

\subsection{Solution to The TDDE in The Dipole Gauge}
In the presence of normally-incident optical terahertz electromagnetic field in the Coulomb gauge, the total Dirac Hamiltonian can be written as: 
\begin{equation}
  H_K=\hbar v_F . \boldsymbol{\sigma} (\mathbf{k}+ \frac{q \mathbf{A}}{\hbar})  
\end{equation}
\begin{equation}
  H_K'=\hbar v_F . \boldsymbol{\sigma} (\mathbf{k'}+ \frac{q \mathbf{A}}{\hbar}) 
\end{equation}
for the $\mathbf{K}$ and  
 $\mathbf{K'}$ points respectively. Here, $\boldsymbol{\sigma}$ is the Pauli matrix defined as $\boldsymbol{\sigma}= \hat{x} \sigma_x + \hat{y}\sigma_y$, the term $(\mathbf{k}+ \frac{q \mathbf{A}}{\hbar})$ is the conical momentum that describes the interaction with the vector potential  $\mathbf{A}(t)$ of the optical light which is  defined as : 
\begin{equation}
\mathbf{A}(t)=\frac{E_0}{\omega_0}e^{-\frac{1}{2} \ln 2(\frac{t}{\tau})^2} [\hat{x} \cos (\frac{\xi}{2}) \cos (\omega_0 t) + \hat{y} \sin (\frac{\xi}{2}) \sin (\omega_0 t)]
\end{equation}
Where $E_0$ is the peak of electrical field strength, $\omega_0 =2 \pi f_0$ is the central frequency, $\tau$ is the pulse width, and $\xi$ is the polarization factor. Solving for the TDDE in the coulomb gauge using the Hamiltonian for the graphene acGNR above, we obtain: 

\begin{equation}
i \hbar \frac{\partial \Psi(\mathbf{r},t)}{\partial t } = \hat H \Psi(\mathbf{r},t)
\end{equation}
Taking into consideration the polarization of the Terahertz field, the solution can expanded as follow:  
\begin{equation}
\Psi(\mathbf{r},t)=\psi^D(\mathbf{r},t)\exp \Big[ \Big( -i \left(\frac{e}{\hbar}\right) \mathbf{r . A}(t) \Big) \Big]
\end{equation} 
Where $ -i \left(\frac{e}{\hbar}\right) \mathbf{r.A}(t)$ is the gauge generation function for transforming the system from Coulomb to dipole gauge. Here, the expansion of the wavefunction $\psi^D(\mathbf{r},t)$ for the four eigenstates of the Hamiltonian
\begin{equation}
\Psi^D(\mathbf{r},t)=c_1 \psi_\mathbf{k}^{(1)} (\mathbf{r}) + c_2 \psi_\mathbf{k}^{(1)} (\mathbf{r})+c_3 \psi_\mathbf{k}^{(1)} (\mathbf{r})+c_4 \psi_\mathbf{k}^{(1)} (\mathbf{r})
\end{equation}
Finally, we obtained the TDDE in this dipole gauge by substituting $\Psi(\mathbf{r},t)$ into Eq. 8 and including the polarization of the incident electric field: 
\begin{equation}
i \hbar \frac{\partial \Psi(\mathbf{r},t)^D}{\partial t }= [\mathbf{H_0}+\hat{V}(t)]\Psi^D (\mathbf{r},t)
\end{equation}
The term $\hat{V}(t)= e \mathbf{E}(t) I_4$ includes the 4x4 identity matrix $I_4$, and $\mathbf{E}(t)$ is the incident electric field. For simplicity, we assumed this electric field to being polarized along the x-axis, resulting in $\hat{V}(t)= q x E_x(t)$ with the polarization factor $\xi = 0$. 
Substituting the wavefunction Eq.10 into Eq.11 and multiplying both sides by $[\mathbf{\Psi_k'}^{(1)}(\mathbf{r})]^\dagger$, yields the equation:
\begin{flalign}
\begin{aligned}
i\hbar(2\pi)^2\Bigg[ \Big (1+ \frac{(k_x - ik_y)(k'_x - ik'_y)}{(k^2+1)(k'^2+1)}-\frac{k' k (ik'_x -ik'_y)}{(k_x - ik_y)}\Big) \dot{c_1}(t) + \\
\Big (1+ \frac{(k_x - ik_y)(k'_x - ik'_y)}{(k^2+1)(k'^2+1)}+\frac{k' k (ik'_x -ik'_y)}{(k_x - ik_y)}\Big) \dot{c_3}(t)]\mathbf{\delta} (\mathbf{k-k'})\\
= e E_x(t) \int d^2\mathbf{r} x e^{i(\mathbf{k-k'}).\mathbf{r}}
\Bigg[ \Big (1+ \frac{(k_x - ik_y)(k'_x - ik'_y)}{(k^2+1)(k'^2+1)}-\frac{k' k (ik'_x -ik'_y)}{(k_x - ik_y)}\Big) c_1(t) + \\
\Big (1+ \frac{(k_x - ik_y)(k'_x - ik'_y)}{(k^2+1)(k'^2+1)}+\frac{k' k (ik'_x -ik'_y)}{(k_x - ik_y)}\Big) c_3(t)\Bigg]
\end{aligned}
\end{flalign}

Repeating the pre-multiplication TDDE of $[\mathbf{\Psi_k'}^{(2)}(\mathbf{r})]^\dagger$, $[\mathbf{\Psi_k'}^{(3)}(\mathbf{r})]^\dagger$, and $[\mathbf{\Psi_k'}^{(4)}(\mathbf{r})]^\dagger$ individually and integrating the the resulting equations over $\mathbf{k'}$ space and using the variable transformation: $c_1(t)=e^{-i\omega_p t}\Tilde{c_1}(t)$, $c_2(t)=e^{-i\omega_p t}\Tilde{c_2}(t)$, $c_3(t)=e^{i\omega_p t}\Tilde{c_3}(t)$, $c_1(t)=e^{i\omega_p t}\Tilde{c_4}(t)$ leads to the four equations of motion corresponding to the four eigenstates of the Hamiltonian system for acGNR: 

\begin{flalign}
\begin{aligned}
    \dot{\Tilde{c_1}}(t) &= \frac{e E_x e^{-i \omega_p t}} {\sqrt{2}\hbar}\Bigg[ \left(\frac{k_x k + (k_x + ik_y)}{k^3} - \frac{k_x -2ik_y}{k_x + ik_y} + \sqrt{\frac{k_x +ik_y}{k}} \right) \Tilde{c_1}(t)\\
     &+\left( \dfrac{k^3(k_x +ik_y)(k_x -2ik_y)}{k_x - ik_y} - \frac{k k_x + k_x + ik_y}{k^3} \right)\Tilde{c_3}(t)\Bigg]&&
\end{aligned}
\end{flalign}

\begin{flalign}
\begin{aligned}
\dot{\Tilde{c_2}}(t) &= \frac{e E_x e^{-\omega_p t}}{\sqrt{2}\hbar} \Bigg[ \bigg( \frac{k k_x + k_x + k_y}{k^3} + \frac{k_x^4 + ik_x^3 k_y + 3k_x k_y^2}{\sqrt{k}} 
\sqrt{(k_x -ik_y)^5 (k_x + ik_y)^3 }
\bigg) \Tilde{c_2}(t) \\
& + \bigg( \frac{k k_x (k_x -ik_y)}{(k_x + ik_y)^2} + 2k_y (k_x -ik_y)^2 + k_x (k_x + ik_y) \bigg) \Tilde{c_4}(t)
\Bigg]&&
\end{aligned}
\end{flalign}

The equations of motion for $\dot{\Tilde{c_3}}(t)$ and $\dot{\Tilde{c_4}}(t)$ are the complex conjugate for the first two, and can be written as follow for simplicity: 

\begin{flalign}
\begin{aligned}
\dot{\Tilde{c_3}}(t)= \big[\dot{\Tilde{c_1}}(t)\big]^*
\end{aligned}
\end{flalign}

\begin{flalign}
\begin{aligned}
\dot{\Tilde{c_4}}(t)= \big[\dot{\Tilde{c_2}}(t)\big]^*
\end{aligned}
\end{flalign}
\subsection{Induced Current Harmonic Spectra}
To calculate the x and y components of the induced-current density, we used the Dirac continuity equation :\\
$\frac{\partial P}{\partial t} + \nabla . \boldsymbol{j}=0$ with $P=q |\Psi(\boldsymbol{r},t)|^2$ being the probability density and $\boldsymbol J$ being the current density. These components of the current density are: 
\begin{equation}
j_x(t)=\frac{q v_0}{(2\pi)^2}[\frac{k_x[\lvert \Tilde{c}_3(t) \rvert^2 + \lvert \Tilde{c}_4(t) \rvert^2 - \lvert \Tilde{c}_1(t) \rvert^2 - \lvert \Tilde{c}_2(t) \rvert^2]}{k} + \frac{i k_y [\Tilde{c}_3 \Tilde{c}^*_1 - \Tilde{c}_4 \Tilde{c}^*_2 -
\Tilde{c}_1 \Tilde{c}^*_3 + \Tilde{c}_2 \Tilde{c}^*_4]}{k}]
\end{equation}

\begin{equation}
j_y(t)=\frac{q v_0}{(2\pi)^2}[\frac{k_y[\lvert \Tilde{c}_3(t) \rvert^2 + \lvert \Tilde{c}_4(t) \rvert^2 - \lvert \Tilde{c}_1(t) \rvert^2 - \lvert \Tilde{c}_2(t) \rvert^2]}{k} + \frac{i k_x [\Tilde{c}_3 \Tilde{c}^*_1 - \Tilde{c}_4 \Tilde{c}^*_2 -
\Tilde{c}_1 \Tilde{c}^*_3 + \Tilde{c}_2 \Tilde{c}^*_4]}{k}]
\end{equation}
Lastly, integrating the current densities over $\mathbf{k}$ gives the total current density, which can be expressed as: $\mathbf{J_q}(t)= g_2 \sum_{\mathbf{k}} \mathbf{j}(t) $ with $g_2=2$ is the degeneracy factor.

\section{Results and Discussion}
Figure 1 shows the net electric current density of the system under excitation conditions. This excitation conditions consist of the electrons movements in the photoexcited state at resonance of $p={\hbar w_0}/{2v_F}$ directional angle of momenta $\phi=\frac{\pi}{4}$, frequency of $f_0=2$ THz, $\tau=1$ ps, and excited by a light pulse of $I=500W/{cm}^2$ on the left figure.
In addition, we conducted the same numerical calculation method and parameters for a different light pulse value $I=1000W/{cm}^2$, which is shown in the right figure.

It is important to note here the current density component $\mathbf{J}_y (t)$ is zero and vanishes when the optical pump is applied, meaning the total current density illustrated in Fig.1 consists solely of $\mathbf{J}_x (t)$, which is due to the x-direction polarized electrical field exposed to the system.  Furthermore, the $I=500W/{cm}^2$ reveals to have a lower current density pump response in comparison to $I=1000W/{cm}^2$ as expected. The peak current density reaches less than $4 \times 10^-{8}(A/cm^2)$ whereas the peak current density reaches almost $0.2(A/cm^2)$ twice for the $I=1000W/{cm}^2$ pump, indicating a difference  within $10^7$ range. Moreover, the two current density noticeable peaks begin at approximately 1700pS, and the second peak for starts at 1900pS. After the mean peaks, the current density does not decay to its initial zero value due to the fact that our model does not include a relaxation term. The lack of the relaxation condition only causes an offset in the current density area but it does not effect the overall conclusion of this study.

\begin{figure}[hbt!]
\includegraphics[width=\textwidth]
{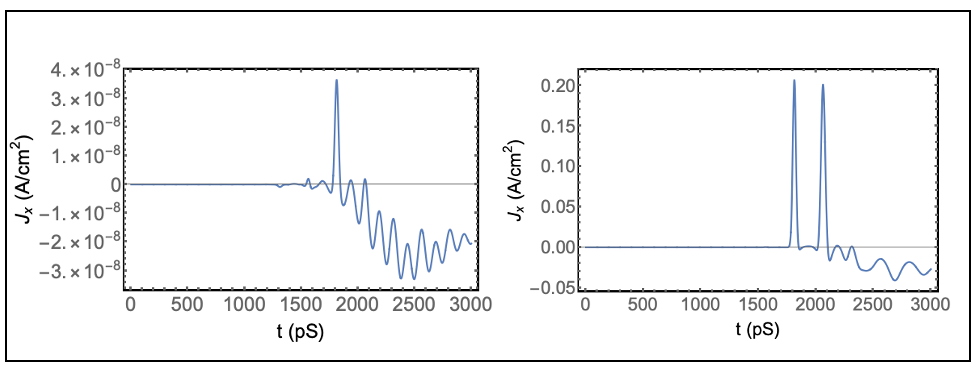}
\caption{The total current density $\mathbf{J}$ with peak pump corresponding to I=$500 W/cm^2$ is shown on the left, and the current density corresponding to I=$1000W/cm^2$ is shown on the right. In both, The resonance of electrons is set for $p={\hbar w_0}/{2v_F}$, the directional angle of momenta is $\phi=\frac{\pi}{4}$, the frequency is $f_0=2$ THz, and the pulse is set to $\tau=1$ ps}
\end{figure} 
To demonstrate the harmonic spectra effect generated by the Rabi Oscillations, we calculated the power of net current density in metallic acGNR Astronomical Unit at the same two intensities as the previous figure.
\begin{figure}[hbt!]
\includegraphics[width=\textwidth]{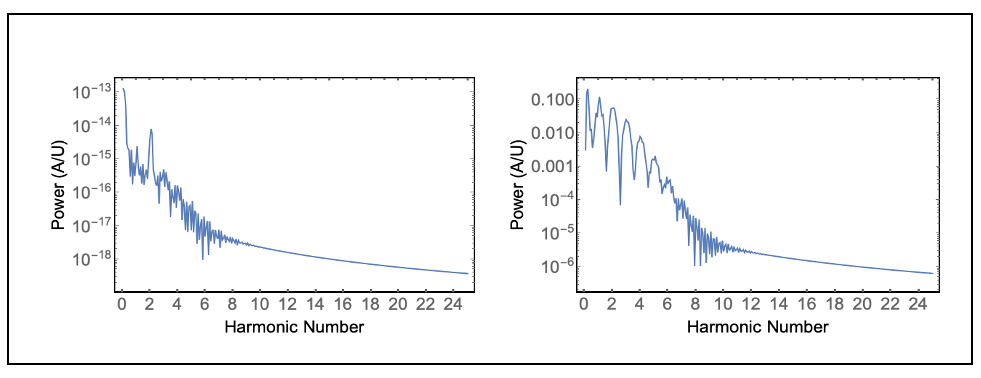}
\caption{Spectral power configuration of the total current density with I =$500W/cm^2$ light pulse in the left figure and I=$1000W/cm^2$ in the right figure. The parameters used in calculating these spectrum is set the same as Fig.1}
\end{figure}

The corresponding spectra is shown is Figure.2 where the power spectra on the left represents I=$500W/cm^2$, here the first four harmonics are distinguishable and the later peaks appear to be fluctuation with no central peak to distinguish whether they are odd or even harmonics.These power fluctuations are continuous up to the tenth harmonic number, then the power decays until it eventually vanishes, Mikhailov [12] described this decay as ${J}\sim 1/m$, where m is the odd harmonics frequencies. However, the first four are odd harmonics surrounded by bifurcations that seem to be at lower frequencies than the main harmonic peak.   
In addition, the result of the harmonic spectra corresponding to I=$1000W/cm^2$ is shown on the right in Fig.2. Here, seven odd harmonic peaks are continuous and clearly distinguishable from the rest of the smaller harmonics and bifurcations around each harmonic does not have a majorly difference frequency than the main harmonic frequency peak. Unlike the plot on the left, the bifurcations in the power plot here is only two until the sixth harmonic number, then they follow the same fluctuations pattern and tend to decay after the 12$^{th}$ harmonic number.

\section{Conclusion}
In conclusion, we used the transformed time-dependent Dirac equation in the dipole gauge describing the unperturbed Hamiltonian of fermions in metallic acGNR to study its nonlinear optical response.  We have analyzed the behaviour of the system with a quantum mathematical model beyond the Boltzman theory when exposed to a strong Terahertz electrical field with difference intensity ranges.  The resultant harmonic spectra has shown to be continuous fluctuations of power with variations of odd and even harmonics. In addition, the peaks decay rapidly with the increase of time and harmonic number as expected in previous studies.  This results confirms previous work that when taking into account the Rabi Oscillations of analyzing the optical response of Graphene structure, the power spectrum behaviour is not limited to odd harmonic.

%\section{References}
%\bibliographystyle{plain}
%\bibliography{ref.bib}

\end{document}